\documentclass[manuscript,authorversion]{acmart} % For ArXiv submission

\AtBeginDocument{%
  \providecommand\BibTeX{{%
    \normalfont B\kern-0.5em{\scshape i\kern-0.25em b}\kern-0.8em\TeX}}}

\acmConference[]{}{}{}
\acmBooktitle{ACM Conference}
\acmPrice{}
\acmISBN{}

\setcopyright{none}
\renewcommand\footnotetextcopyrightpermission[1]{}

\usepackage{graphicx}
\usepackage{subfig}
\usepackage{booktabs}% http://ctan.org/pkg/booktabs

\usepackage{color,soul}

\newcommand{\anon}[1]{#1}  % not hiding anonymity

\usepackage{enumitem}
\newlist{RQ}{enumerate}{1}
\setlist[RQ]{label=RQ\arabic*:}
\usepackage{tabularx}
\usepackage[normalem]{ulem}
\usepackage{comment}

\usepackage{xparse}

\NewDocumentCommand{\showTweetId}{m}{%
  \href{https://twitter.com/user/status/#1}{#1}  % a space follows
}

\NewDocumentCommand{\citeTweet}{ >{\SplitList{,}} m }{%
  \footnote{
    Tweet IDs=\ProcessList{#1}{\showTweetId}. Accessible via \href{https://developer.twitter.com/en/docs/twitter-api/tweets/lookup/quick-start}{Twitter API}.
  }
}

\begin{document}

\title[Image Cropping on Twitter]{Image Cropping on Twitter: Fairness Metrics, their Limitations, and the Importance of Representation, Design, and Agency}

\author{Kyra Yee}
\authornote{Authors contributed equally to this research.}
\email{kyray@twitter.com}
\author{Uthaipon Tantipongpipat}
\authornotemark[1]
\email{uthaipon@gmail.com}
\author{Shubhanshu Mishra}
\authornotemark[1]
\email{smishra@twitter.com}
\affiliation{%
  \institution{Twitter}
  \streetaddress{1355 Market Street Suite 900}
  \city{San Francisco}
  \state{California}
  \country{USA}
  \postcode{94103}
}

\renewcommand{\shortauthors}{Kyra Yee, Uthaipon Tantipongpipat, \& Shubhanshu Mishra}
\newcommand{\cut}[1]{}

\begin{abstract}

Twitter uses machine learning to crop images, where crops are centered around the part predicted to be the most salient. In fall 2020, Twitter users raised concerns that the automated image cropping system on Twitter favored light-skinned over dark-skinned individuals, as well as concerns that the system favored cropping woman's bodies instead of their heads. In order to address these concerns, we conduct an extensive analysis using formalized group fairness metrics. We find systematic disparities in cropping and identify contributing factors, 
including the fact that the cropping based on the single most salient point can amplify the disparities because of an effect we term \textit{argmax bias}.
However, we demonstrate that formalized fairness metrics and quantitative analysis on their own are insufficient for capturing the risk of representational harm in automatic cropping. We suggest the removal of saliency-based cropping in favor of a solution that better preserves user agency. For developing a new solution that sufficiently address concerns related to representational harm, our critique motivates a combination of quantitative and qualitative methods that include human-centered design.

\end{abstract}




\maketitle

\section{Introduction} \label{sec:intro}

Automated image cropping (or smart image cropping) \cite{adobesmartcrop} is a task to, given a viewport dimension or aspect ratio and an image, crop the image such that the image fits the viewport or aspect ratio (\textit{width/height}) while ensuring that its most relevant or interesting parts are within the viewport. The idea of automated image cropping \cite{chen2016automatic, chen2017quantitative} has been present in the industry since at least 1997 \cite{bollman1999automatic}.

Image cropping have had applications in several fields. It has a rich history in the field of cinematography and broadcasting, where film footages's aspect ratio is changed to match the display aspect ratio. Cropping an image to show its most interesting part is useful and often done manually in artistic settings. In recent website design, image crop is used to develop approaches for responsive images which do not distort the aesthetics of the website or app \cite{twittersaliencyblog}.  This is often used with responsive images where a pre-defined image on a website or app needs to be resized to fit devices with different dimensions \cite{mozilla_image}.

The modern automation of cropping process is especially useful for many reasons. First, automation significantly reduces human burden when the throughput of images to be processed is very high. For example, automated image cropping is used to show user submitted images, which are very large in size, by many platforms such as Facebook, Instagram, Twitter, and Adobe, in conjunction with various degrees of user controls \cite{instagramcrop, adobesmartcrop, twittersaliencyblog,facebookvideocrop}. Second, many modern platforms are operating on multiple types of devices requiring varying aspect ratios, increasing the number of crops needed even on a single image. Again, automation reduces human burden in the multiplicities of cropping required. In addition, automated image cropping can help individuals identify important content in the image and do it faster compared to no-crop \cite{Bongwon2003}. Automated image crops can also produce more desirable thumbnails to users compared to shrinking the whole image to fit the viewport \cite{Xie2006BrowsingLP}.

However, automating image cropping can also result in errors and other undesirable outcomes. Often, automation is implemented by machine learning (ML) systems which are designed to ensure that the average error is low, but do not account for the disparate impact of cases where the error is not uniformly distributed across demographic groups \cite{buolamwini2018,objectrecbias}. Additionally, there have been many calls for a more critical evaluation of the underlying values and assumptions embedded in machine learning and human computer interaction (HCI) systems \cite{dotan2019value,bardzell, borning2012next, friedman2002value, friedman2008value, van2011feminist}.  

\begin{figure}
    \centering
    \includegraphics[width=\textwidth]{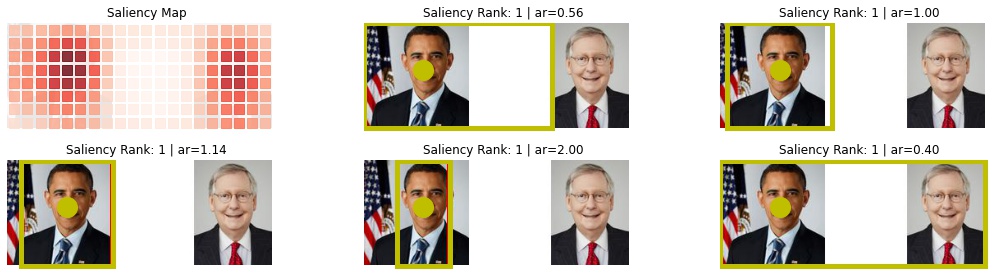}
    \caption{Automated image crops (yellow rectangles) on the image of two individuals separated by the white background in the middle. Crops are produced by Twitter saliency-based model for different aspect ratios. \textbf{Top left:} the heat map of saliency scores produced by Twitter image cropping model on the image. Locations with high predicted saliency could be humans, objects, texts, and high-contrast backgrounds. \textbf{Others:} the original image with its maximum saliency point (as the yellow dot) and the automatic crop (as the yellow rectangle) with different aspect ratios (the `ar' values).
    Images from Wikidata.
    }
    \label{fig:obama-mitch}
\end{figure}

Our work focuses on one automated image cropping system, namely Twitter's saliency-based image cropping model, which automatically crops images that users submit on Twitter to show image previews of different aspect ratios across multiple devices. The system employs a supervised machine learning model on existing saliency maps to predict saliency scores over any given image (as an input). Saliency scores are meant to capture the ``importance'' of each region of the image. After having the saliency scores, the model selects the crop by trying to center it around the most salient point, with some shifting to stay within the original image dimension as needed. See Figure \ref{fig:obama-mitch} for an example. While Twitter platform enjoys several benefits of automation of image cropping as mentioned earlier, several concerns exist around its usage, namely concerns around \textit{representational harm}. Drawing on the work of \cite{collins2002black, crawford2017trouble}, we define representational harm as the harm associated with a depiction that reinforces the subordination of some groups along the lines of identity, such as race, class, etc., or the intersection of multiple identities \cite{crenshaw1989demarginalizing}. Types of representational harm include stereotyping, failure to recognize someone's humanity, under-representation, or denigration of an individual or group of people \cite{crawford2017trouble}. This work is meant to directly address the concerns that Twitter users themselves have raised on the platform. We summarize the concerns as follows:
\begin{enumerate}
    \item \textbf{Unequal treatment on different demographics.} In September 2020, a series of public Tweets claimed that Twitter image cropping crops to lighter-skinned individuals when lighter and darker-skinned individuals are presented in one image\citeTweet{1307115534383710208, 1307427332597059584, 1307440596668182528}
    (see Figure \ref{fig:obama-mitch} as an example of such image),
    spurring further online responses
    \cite{twitterresponse} \citeTweet{1307777142034374657}, news \cite{guardiannews}, and articles \cite{davidarticle}. However, as one user noted \citeTweet{1307427207489310721}, one limitation of people posting individual Tweets to test cropping behavior is that giving one specific example is not enough to conclude there is systematic disparate impact. In order to address this, a Twitter user performed an independent experiment \cite{vinayarticle} of 92 trials comparing lighter and darker-skinned individuals using the Chicago Faces Dataset \cite{chicagoface}.
    Because of the small size of experiment, there can still be concerns today whether the Twitter model systematically favors lighter-skinned individuals over darker-skinned, or more generally over other demographics such as men over women.

    \item \textbf{Male gaze.} Another concern arises when cropping emphasizes a woman's body instead of the head.\citeTweet{1010288281769082881}Such mistakes can be interpreted as an algorithmic version of \emph{male gaze}, a term used for the pervasive depiction of women as sexual objects for the pleasure of and from the perspective heterosexual men \cite{mulvey1989visual,korsmeyer2004feminist}.
    
    \item \textbf{Lack of user agency.} The automation of cropping does not include user agency in dictating how images should be cropped. This has the potential to cause representational harm to the user or the person present in the photo, since the resulting crop may change the interpretation of a user's post in a way that the user did not intend.
    
\end{enumerate}

\paragraph{Research Questions}
The questions we try to answer in this work are motivated from the above concerns, and can be summarized as follows:
\begin{RQ}
    \item 
    To what extent, if any, does Twitter's image cropping have disparate impact (i.e. systematically favor cropping) people on racial or gendered lines?
    \label{Ques:bias_exist}
    
    \item What are some of the factors that may cause systematic disparate impact of the Twitter image cropping model? 
    
    \item What are other considerations (besides systematic disparate impacts) in designing image cropping on platforms such as Twitter? What are the limitations of using quantitative analysis and formalized fairness metrics for surfacing potential harms against marginalized communities? Can they sufficiently address concerns of representational harm, such as male gaze? 
    \label{Ques:male_gaze}
    
   \label{Ques:explain_bias}
    
    \item What are some of the alternatives to saliency-based image cropping that provide good user experience and are socially responsible? How do we evaluate such alternatives in a way that minimizes potential harms? \label{Ques:solutions}
\end{RQ}

RQ1 and RQ2 arise naturally from the unequal treatment concern from users. However, some concerns such as the lack of user agency are not adequately captured by quantitative fairness metrics, leading to RQ3. Finally, RQ4 aims to apply the analysis and discussions to suggest and evaluate other image cropping solutions. 

\paragraph{Summary of Contribution}
Our contribution can be summarized as follows, following the research questions in their order:
\begin{enumerate}
    \item We provide quantitative fairness analysis showing the level of disparate impact of Twitter model over racial and gender subgroups.\footnote{We note the limitations of using race and gender labels, including that those labels can be too limiting to how a person wants to be represented and do not capture the nuances of race and gender. We discuss the limitations more extensively in Section \ref{sec:limit_and_future_work}.}
    We also perform additional variants of experiments and an experiment on images showing the full human body to gain insights into causes of disparate impact and male gaze.
    \item Using the results of and observations from the fairness analysis, we give several potential contributing factors to disparate impact. Notably, selecting an output based on a single point with highest predicted scores (the \textit{argmax selection}) can amplify disparate impact in predictions, not only in automated image cropping but also in machine learning in general.
    \item We qualitatively evaluate automated image cropping, discussing concerns including user agency and representational harm. We argue that using formalized group fairness metrics is insufficient for surfacing concerns related to representational harm for automated image cropping.
    \item We give alternative solutions, with a brief discussion of their pros and cons and how they can generalize to other settings beyond Twitter image cropping model. We emphasize the importance of representation, design, user control, and a combination of quantitative and qualitative methods for assessing potential harms of automated image cropping and evaluating alternative solutions.

\end{enumerate}

\paragraph{Paper Organization}
%The paper is organized as follows. 
In Section \ref{sec:related_work}, we give background and related work on the Twitter saliency-based image cropping and fairness and representational harm in machine learning. We answer RQ1 and provide the first contribution in Section \ref{sec:dataset}-\ref{male gaze section}, including the methodology of the experiments. We answer RQ2 and provide the second contribution in Section \ref{sec:argmax}-\ref{sec:other_explanation_bias}. We answer RQ3 and RQ4 and provide the third and fourth contributions in Section \ref{sec:qualitativecriticism} and \ref{sec:solutions}, respectively.
We list limitations and future directions in Section \ref{sec:limit_and_future_work} and conclude our work and contribution in Section \ref{sec:conclusion}.

\section{Background and Related Work} \label{sec:related_work}
\subsection{On Ethical Human-Computer Interaction and Machine Learning within Industry}
\paragraph{The Problem of Ethical Design}
Algorithmic decision making and machine learning have the potential to perpetuate and amplify racism, sexism, and societal inequality more broadly. Design decisions pose unique challenges in this regard since design ``forges both pathways and boundaries in its instrumental and cultural use'' \cite{noble}. 
Examples of challenges in designing ethical ML systems include:\footnote{This list of challenges is non-exhaustive. Notably, there is a broad body of work on challenges related to AI ethics in industry, including ethics-washing, regulation, corporate structure and incentives, and power not included in the discussion here. 
} 
\begin{enumerate}
    \item Bridging human and ML perspectives: Because ML systems can make seemingly nonsensical errors that lack common sense because they can pick up on spurious correlations, it may be hard for designers to bridge the ML and the human perspective, and to anticipate problems \cite{dove2017ux}.
    \item Dataset: For image processing, one key challenge in validating and communicating such problems is the lack of high quality datasets for fairness analysis available \cite{buolamwini2018}, especially to industry practitioners \cite{andrus2021we}.
    \item Lack of universal formalized notion of fairness : One key focus of ethical machine learning has been on developing formal notions of fairness and quantifying algorithmic bias. Another challenge is the lack of a universal formalized notion of fairness that can be easily applied to machine learning models; rather, different fairness metrics imply different normative values and have different appropriate use cases and limitations \cite{barocas-hardt-narayanan,narayanan2018translation}. 
    \item Lack of a satisfactory definition of universally appropriate metrics or optimization objectives for machine learning for certain classes of problems: The challenge of a lack of a universally "correct" answer or value system applies to metrics used in machine learning beyond fairness as well. \citet{vallor_2021, tasioulas_2021}  argue machine learning is not always suitable for all problems because a fundamental risk in its deployment is the necessity of defining a universally "optimal" state when such a thing does not exist and is not appropriate for many classes of problems. For example, \citet{vallor_2021} notes the absurdity of trying to optimise color for the optimal painting.
\end{enumerate} 

\paragraph{Prior Work in Ethical Design}
Prior frameworks integrating ethics and design includes value sensitive design, which advocates for a combination of conceptual, empirical, and technical analysis to ensure social and technical questions are properly integrated \cite{borning2012next, friedman2002value, friedman2008value}. Similarly, feminist HCI methods advocate for a simultaneous commitment to moral and scientific agendas, connection to critical feminist scholarship, and co-construction of research agendas and value-laden goals \cite{bardzell}.
Another commonly used framework is human-centered design, which aims to center human dignity, rights, needs, and creativity \cite{buchanan2001human, gasson2003human, gill1990summary}. Human-centered design arose out of a critique to user centered approaches, which \citet{gasson2003human} argues fails to question the normative assumptions of technology because of a ``goal directed focus on the closure of predetermined, technical problems''. \citet{costanza2018design}'s design justice builds on previous frameworks but emphasizes community led design and the importance of intersectionality and participatory methods for conceptualizing user needs across different marginalized identities.
In critiqueing how design practices have historically erased indigenous epistemologies and ways of being, \citet{escobar2018designs} calls for design practices that accomodate a plurality of values, which \cite{costanza2018design} notes even human-centered or participatory approaches often fail to accomplish. 

\paragraph{How This Work Contributes}
This work is meant to contribute to a broader conversation about how to promote ethical human-computer interaction within the technology industry, which will require open communication between industry, academia, government, and the wider public to solve, as well as an acknowledgment of the responsibility of companies to be held accountable for the societal effects of their products. In order to promote transparency and accountability to users, we strive to create a partnership between social media platforms and users where users interface with social media while maintaining control of their content and identities on the platform. We also recognize it is critical to attend to the perspectives and experiences of marginalized communities not only because “it empowers a comparatively powerless population to participate in processes of social control, but it is also good science, because it introduces the potential for empirically derived insights harder to acquire by other means” \cite{bardzell}. In particular, the research questions in this work sprang up from the concerns for and reported from users, and our recommendations (Section \ref{sec:solutions}) are evaluated through the lens of ethical human-centered design, which motivates our focus on user agency and the effects of the algorithm on marginalized communities.
This work is not meant to solve all the challenges in ethical design, but rather to underscore the importance and utility of incorporating ethical design principles when evaluating the societal affects of machine learning in real world use cases, such as saliency based cropping. 

\subsection{History of Representational Harm in Technology}
\label{sec:historyrepresentationalharm}
\citet{crawford2017trouble} details how addressing representational harm has been historically under-addressed in the machine learning community, and provides a taxonomy of representational harms, including stereotyping, denigration, recognition, ex-nomination, and under representation. Here, we provide some background and examples of the harm.

\paragraph{Representational Harm and Feminist and Critical Race Studies}
Challenging stereotypical and harmful representations of marginalized groups has been a core theme in feminist and critical race studies. \citet{collins2002black} describes how harmful dominant representations such as mammies, welfare recipients, and the Jezebel are "controlling images" that are used to continually reproduce and entrench the intersecting oppression of Black women. 
She suggests that self-representation is an essential tool to dismantling dominant representations, a strategy that has been called for across many different feminist camps \cite{nash2008strange}.
However, \citet{nash2008strange} argues that self-representation on its own has significant limitations given that images are seen through a ``racially saturated field of visibility'', where the production of the visible and the seen occurs on racialized lines \cite{butler2020endangered}. \citet{butler2020endangered} originally coined this phrase to describe how video of Rodney King being brutally beaten was violently recontextualized via selective editing as well as racist confirmation bias of jurors to serve as evidence against him \cite{butler2020endangered}. 

\paragraph{Representational Harm and Unequal Visibility in Technology}
 \citet{benjamin} introduces the notion of coded exposure, which describes how people of color suffer simultaneously from under and over exposure in technology. In her work, she demonstrates “some technologies fail to see blackness, while other technologies render Black people hyper visible” \cite{benjamin}. Beginning with the development of color photography, Shirley cards used by Kodak to standardize film exposure methods were taken of white women, causing photos of darker-skinned people to be underexposed and not show up clearly in photos \cite{benjamin}.  
More recently, \citet{buolamwini2018} demonstrated that widely used commercial facial recognition technology is significantly less accurate for dark-skinned females than for light-skinned males. Simultaneously, in the United States, facial recognition has rendered communities of color hyper visible since it is routinely used to surveil communities of color \cite{benjamin, devich2020defund}. Similarly, facial recognition is being used in China to surveil Uighurs as part of a broader project of Uighur internment, cultural erasure, and genocide \cite{uighurprofile}, and facial recognition tools have been specifically developed to identify Uighurs \cite{ethnicitydetection}. 
\citet{objectrecbias} demonstrates that images from ImageNet, the dataset widely used across a range of image processing tasks, come from primarily Western and wealthy countries, which is a contributing factor for why commercial object recognition systems performs poorly for items from low income and non-Western households. This ``unequal visibility'' is not just limited to images but also texts, e.g. \citet{Mishra2020} report that names from certain demographics are less likely to be identified as person names by major text processing systems.

\paragraph{The Urgency to Address Representational Harm}
These examples illustrate how technologies that are presented as objective, scientific, and neutral actually encode and reproduce societal unequal treatment over different demographics, both in their development and deployment. Additionally, they underscore the highly contextual nature of representational harm \cite{crawford2017trouble}; not all exposure is positive (as in the case of surveillance technology or stereotyping, for example), and representational harm provides a unique challenge to marginalized communities who have faced repeated challenges to maintaining their privacy and in advocating for positive representations in the media.
Although representational harm is difficult to formalize due to its cultural specificity, it is crucial to address since it is commonly the root of disparate impact in resource allocation \cite{crawford2017trouble, collins2002black}. For instance, ads on search results of names perceived as Black are more likely to yield results about arrest records, which can affect people's ability to secure a job \cite{crawford2017trouble,sweeney2013discrimination}.  Although allocative harm may be easier to quantify than representational harm, it is critical that the ML ethics community works to understand and address issues of representational harm as well \cite{crawford2017trouble}. 

\paragraph{How This Work Contributes}
This work views the problem of automatic image cropping through the lens of representational harm, with the goal of not reproducing or reinforcing systems of subordination for marginalized communities in automatic image cropping. 

Specifically, this work includes the qualitative analysis (Section \ref{sec:fairness_results}) to measure the extent that the image cropping model misrepresents certain race and gender as more ``salient'' (i.e. worthy of focus and being the center of a crop), discusses the fundamental risk representational harm for saliency based image cropping (Section \ref{sec:qualitativecriticism}), and provides recommendations with representational harm concern in perspective (Section \ref{sec:solutions}).

\subsection{Background on Twitter Saliency Image Cropping Model}
Twitter's image cropping algorithm \cite{twittersaliencyblog} relies on a machine learning model trained to predict \textit{saliency}. Saliency is the extent that humans tend to gaze on a given area of the image, and it was proposed as a good proxy to identify the most important part of the image to center a crop around as importance is relative and difficult to quantify \cite{saliencycrop}. Locations with high predicted saliency (or \textit{salient regions}) include humans, objects, texts, and high-contrast backgrounds (see examples in \cite{huang2015salicon} and Figures \ref{fig:obama-mitch},\ref{fig:male-gaze-chest},\ref{fig:multimodal-crops} for Twitter models). The Twitter algorithm finds the most salient point, and then a set of heuristics is used to create a suitable center crop around that point for a given aspect ratio. We note that other automated cropping algorithms exist which utilize the full saliency map on an image to identify the best crop, such as by covering the maximum number of salient points or maximizing the sum of saliency scores \cite{DBLP:journals/corr/ChenHCTCC17}. 

Saliency prediction can be implemented by deep learning, a recent technique that has significantly increased prediction performance, though there are still gaps in prediction compared to humans and limitations such as robustness to noise \cite{borji2018saliency}.
Twitter's saliency model builds on the deep learning architecture DeepGaze II \cite{deepgazeii}, which leverages feature maps pretrained on object recognition. In order to reduce the computational cost to be able to use the model in production, Twitter leverages a combination of fisher pruning \cite{fisherpruning} and knowledge distillation \cite{hinton2015distilling}. The model is trained on three publicly available external datasets: \citet{salicon,MIT1003,cat2000}. 
See \citet{fisherpruning} for more precise details on model architecture.

Cropping is conducted as follows:
\begin{enumerate}
    \item For a given image, the image is discretized into a grid of points, and each grid point is associated with a saliency score predicted by the model. 
    \item The image along with the coordinates of the most salient point and a desired aspect ratio are passed as an input to a cropping algorithm. This is repeated for each aspect ratio to show the image on multiple devices. 
    \item If the saliency map is almost symmetric horizontally\footnote{We rescale the image to a 10x10 grid and use the 95th-percentile of the absolute difference in saliency scores between the left and the right of the rescaled image. We check whether the 95th-percentile is below a certain threshold after some normalization.} For exact details, please see the open-source code at \anon{ \url{https://github.com/twitter-research/image-crop-analysis/blob/main/src/crop_api.py}}, 
    then a center crop is performed irrespective of the aspect ratio. 
    \item Otherwise, the cropping algorithm tries to ensure that the most salient point, or the \textit{focal point}, is within the crop with the desired aspect ratio. This is done by cropping only one dimension (either width or height) to achieve the desired aspect ratio. 
\end{enumerate}

\section{Quantitative Analysis and Results of Image Cropping Using Group Fairness Metrics} \label{sec:fairness_results}

In this section, we answer the first research question on whether Twitter's image cropping has disparate impact on racial or gendered lines by performing an analysis measuring disparate impact based on the fairness notion of \textit{demographic parity}. We first define demographic parity, and then describe the dataset, experiment methodology, and results.

\paragraph{Demographic Parity}
Automatic image cropping was developed so that the most interesting and noticeable part of an image is displayed for users to see. However, in an image with multiple people, sometimes it is impossible to find a crop that will include all people that will fit the desired aspect ratio. This puts the model in a difficult position since there is no “ideal” solution. In handling these cases, what is a fair way to determine who should be cropped in and out? One central concern is that the model should not be systematically cropping out any demographic group such as dark-skinned individuals. Although many competing definitions of fairness have been proposed \cite{barocas-hardt-narayanan, narayanan2018translation}, one notion of fairness that may seem suitable for this context is demographic parity (also referred to as independence or statistical parity). The intuition behind using demographic parity is that the model should not be favoring representing one demographic group over another, so in cases where the model is forced to choose between two individuals, the rate at which they are cropped out should be roughly equal. 
Another interpretation is that roughly equal rates of cropping mitigates the representational harm associated with under-representation \cite{crawford2017trouble}.
For our purposes, given a set of images each depicting two people from group $a$ and group $b$, the auto-cropping model has disparate impact (or a lack of demographic parity for group $a$ with respect to group $b$)  if 

$$
\frac{P(R=1|A=a)}{P(R=1|A=b)} \leq 1 - \epsilon
$$
where $R=1$ denotes that the most salient point (the focal point for the crop) is on the person, $A$ represents group status \cite{feldman2015, barocas-hardt-narayanan}, $\epsilon$ is a slack value, and $P(R=1|A=a)$ is the probability that when two persons, one from $a$ and one from $b$, are given, the auto-cropping model has the most salience on the person from $a$.  Note that we may swap $a$ and $b$ to ensure representation for the class $a$ as well. 

\subsection{Dataset} \label{sec:dataset}

We provide our detailed methodology for using the wikidata API to recreate the dataset with our suggested list of filters. Since our methodology works with any dataset which has the required categorical information about each image, the reader should be able to easily replicate this approach on any new data. We use Wikidata Query Service \cite{wikiquery} to create our WikiCeleb dataset\footnote{
The code of the work is open-sourced at \anon{ \url{https://github.com/twitter-research/image-crop-analysis}}. The WikiCeleb dataset we use in this paper was collected through this methodology on November 2020.} consisting of images and labels of celebrities (individuals who have an ID in the Library of Congress names catalogue\footnote{The Library of Congress Name Authority File (NAF) file provides authoritative data for names of persons: \url{https://id.loc.gov/authorities/names.html}}) in Wikidata. 
Given that ancestry and gender identity of celebrities (especially those identified by Library of Congress Name Authority File) is generally known, using these manually curated labels is less concerning compared to automatically generating sensitive labels by a model which may exacerbate disparity of errors in labels. If an individual has more than one image, gender, or ethnicity, we sample one uniformly at random. We filter out one sensitive occupation\footnote{
Those whose occupation is pornographic actor (Q488111; see \url{https://www.wikidata.org/wiki/Q488111}) are excluded to avoid potentially sensitive images.} and any images that are before the year 1950 to avoid grey-scale images.
We note the limitations of using race and gender labels, including that those labels can be too limiting to how a person wants to be represented and do not capture the nuances of race and gender. We discuss the limitations more extensively in Section \ref{sec:limit_and_future_work}.

\paragraph{WikiCeleb Specifications}
WikiCeleb consists of images of individuals who have Wikipedia pages, obtained through the Wikidata Query Service. It contains 4073 images of individuals labeled along with their gender\footnote{{Property:P21}. \url{https://www.wikidata.org/wiki/Property:P21}} identity and ethnic group\footnote{{Property:P172}. \url{https://www.wikidata.org/wiki/Property:P172}} as recorded on Wikidata. We restrict our query to only consider male\footnote{{Q6581097}. \url{https://www.wikidata.org/wiki/Q6581097}} and female\footnote{{Q6581072}. \url{https://www.wikidata.org/wiki/Q6581072}} gender as the data size for other gender identities is very sparse for our query.\footnote{We recognize that restricting gender to cis-gender woman and men is an extreme simplification of gender, and only use it as a rough approximation here. See \ref{sec:limit_and_future_work} for a more in depth discussion of the limitations.} Ethnic group annotations may be more fine grained than racial groupings, so we map from ancestry to more course racial groupings, as defined by the US census \cite{censusrace} (e.g. Japanese $\rightarrow$ Asian).
We discard ethnicities with small (<40) samples. For small subsets (<5\%) of individuals with multiple labeled ethnicities, we sample one ethnicity uniformly at random. Finally, we split the data into subgroups based on race and gender, and drop subgroups with less than 100 samples. This results in four subgroups: Black-Female, Black-Male, White-Female, and White-Male, of size 621, 1348, 213, and 606, respectively.

\subsection{Methodology}
We perform analysis on all six pairs of subgroups across four subgroups (Black-Female, Black-Male, White-Female, and White-Male). On each pair, we sample one image independently and uniformly at random from each of the two groups and attach those two images horizontally (padding black background when images have different heights).\footnote{We have also performed an experiment by attaching two images vertically and observed several of the saliency map predictions. We found the difference of saliency maps on individual images between horizontal and vertical attaching to be negligible. 
} We run the saliency model on the attached image and determine what image the maximum saliency point lies on. The group from which the image has maximum saliency point is defined to be \textit{favored} (by the saliency model compared to the other group) in this sampling pair. We repeat the process on this pair of groups 10000 times, and record the ratio of times (or probability) that the first group is favored.

\subsection{Results}
\begin{figure}[ht] 
  \centering
  \includegraphics[width=0.7\linewidth]{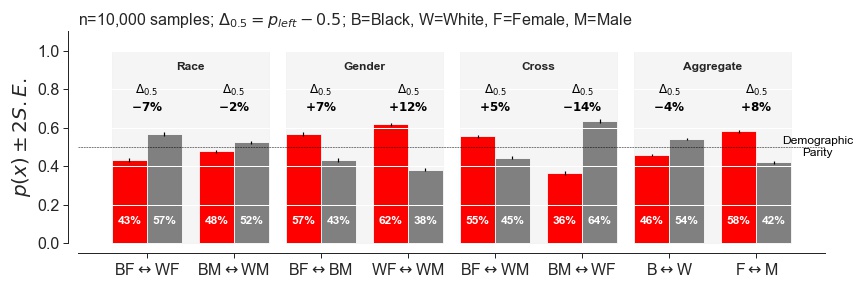}
\includegraphics[width=0.29\linewidth]{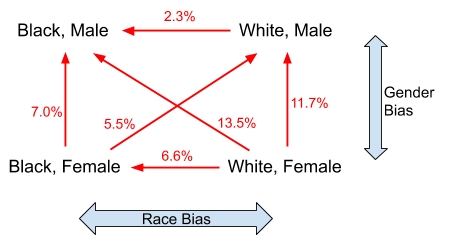}
  \caption{\textbf{Left}: The proportion of times (on y-axis) a group was selected as the crop focus when comparing each pair of groups (on x-axis). The right most panel aggregates subgroups over race or gender. \\\textbf{Right}: Same as left figure, but in a graph format. Four vertices represent four subgroups, and each arrow is one comparison. The head of the arrow is the less favored group. The x\% indicates x\% deviation from 50\%. For example, White female images are more favored than Black male images with probability 0.635. Each has $\pm1.0\%$ (after rounding) with a 95\% confidence interval.}
  \label{fig:base_result}
\end{figure}

For each of the six pairs of subgroups, we report the probability that the first group is favored. The results are in Figure \ref{fig:base_result}, also represented differently to show race and gender disparate impact more clearly. We observe the model's strong gender favoritism for females over males, and a smaller favor for white over Black. The model's gender and race favoritism is stronger among favored intersecting subgroups (white or female) than the other (Black or male). The 95\% confidence intervals of $\pm1.0\%$ are from random sampling and imply that probabilities of favored groups beyond 51\% are statistically significant sign of the model favoring one demographic over another in the dataset.

\begin{figure}[ht] 
\begin{minipage}{.4\textwidth} 
  \centering
  \includegraphics[width=0.8\linewidth]{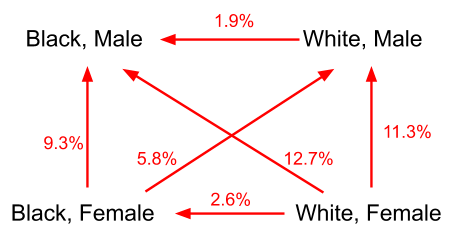}
  \caption{We scale the images to a fixed height before performing the experiment. The sample size is 5000, and each probability has $\pm1.4\%$ (after rounding) as its 95\% confidence interval. We see no significant difference in results from the original in Figure \ref{fig:base_result} (\textbf{Right}).} 
  \label{fig:result_scale}
\end{minipage}
\ \ \
\begin{minipage}{.53\textwidth}
  \centering
  \includegraphics[width=0.6\linewidth]{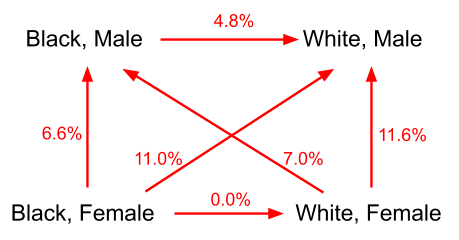}
  \caption{We find maximum of saliency score of each image individually before comparing them, instead of attaching images first then find the max saliency point. The probability is computed over all possible image pairs on each subgroup comparison. Model disparity on race, but not on gender, has significantly changed compared to Figure \ref{fig:base_result} (\textbf{Right}).}
  \label{fig:result_no_attach}
\end{minipage}
\end{figure}

In addition, we perform variants of the experiments as follows.
\paragraph{Scaling of Images}
Images in the dataset vary in sizes. In this experiment variant, we scale all images to a fixed height of 256 pixels while fixing their aspect ratios. The results are similar as shown in Figure \ref{fig:result_scale}, showing that scaling of images has no significant effect.

\paragraph{Attaching Images}
One question we may ask is whether the disparate impact in saliency prediction is already intrinsic to the images in itself, or is contributed by attaching them together. In this variant, we run the saliency model on individual images first, then sample one image from each subgroup in the pair, and select the image with higher maximum saliency score. The results are in Figure \ref{fig:result_no_attach}, showing that attaching images indeed changes disparate impact across race but not gender. In particular, the model's original race favoritism for White when attaching images in the male subgroup is flipped to favoring Black when images are not attached, and in Female group the favoritism for white is diminished.

\subsection{Male Gaze} \label{male gaze section}

In this section, we discuss the male gaze concern, i.e. the claim that saliency-based models are likely to pay attention to woman's body or exposed skin. Male gaze can be thought of as a stereotyping harm, since representations imbued with male gaze reinforce stereotypes of woman's women’s role as sexual objects available for men’s consumption \cite{mackinnon1987feminism} .

In order to study this assumption, we randomly selected 100 images per gender from the WikiCeleb dataset which 1) have height/width ratio of at least 1.25 (this increases the likelihood of image containing full body), and 2) have more than one salient region (salient regions are distinctive regions of high saliency identified by segmenting the saliency map\footnote{Salient regions were identified in the saliency map using the regionprops algorithm in the scikit-image library \url{https://scikit-image.org/docs/dev/auto_examples/segmentation/plot_label.html}}. This increases the likelihood of the image having a salient point other than the head). 

\begin{figure}
    \centering
    \includegraphics[width=0.4\textwidth]{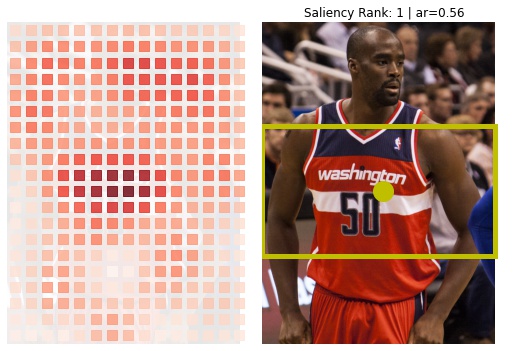}
    \includegraphics[width=0.4\textwidth]{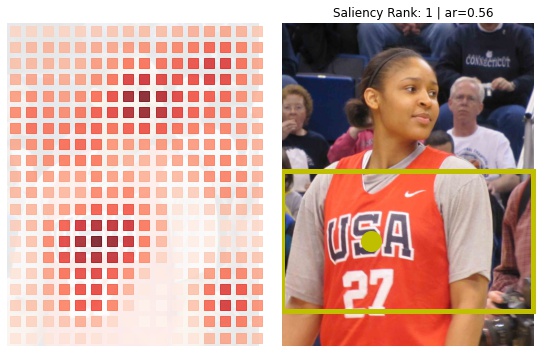}
    \\
    \includegraphics[width=0.4\textwidth]{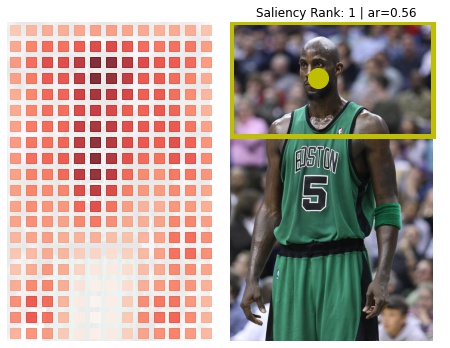}
    \includegraphics[width=0.4\textwidth]{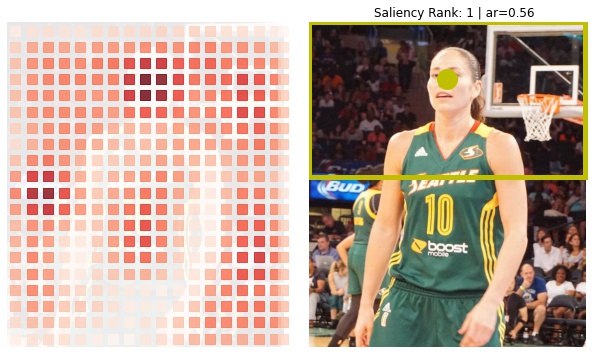}
    \caption{Body areas can be salient because of certain artifacts in images. \textbf{Top:} Images where the crop is on the body because of jersey numbers.  \textbf{Bottom:} similar images where the crop is on the head but jersey number is among a major salient point. Images used in this figure are of public figures available on Wikidata.}. 
    \label{fig:male-gaze-chest}
\end{figure}

We found that no more than 3 out of 100 images per gender have the crop not on the head. The crops not on heads were due to high predicted salient scores on parts of the image such as a number on the jersey of sports players or a badge. These patterns were consistent across genders. For example, see Figure \ref{fig:male-gaze-chest} where the crop is positioned on the body of the individual, and the closeups of the salient regions confirm that the crop is focused on the jersey number while the head is still a significant salient region. Our analysis also replicates when we used smaller (10 for each gender) targeted samples depicting exposed skin (e.g. sleeveless tops in males and female; bare chest for males and low-cut tops for females). Our findings again reveal high saliency on head or on jersey of individuals (see bottom figures in Figure \ref{fig:male-gaze-chest}). Finally, we spot checked the same images via another publicly available model from the Gradio App \cite{kroner2020contextual} (this model crops based on the largest salient region), here too the results replicate. A more rigorous large-scale analysis can be conducted to ascertain the effect size, but this preliminary analysis reveals that the models we studied may not explicitly encode male gaze as part of their saliency prediction and that the perception of male gaze might be because of the model picking on other salient parts of the image like jersey numbers (or other text) and badges which are likely to be present in the body area. Hence, the crops likely confound with the body-focused crop in the rare cases that it happens. 

\subsection{Argmax Selection Amplifies Disparate Impact \textit{: Argmax Bias}} \label{sec:argmax}

We consider the additional impact of ML models which use the argmax approach for inference when applied to social settings. This concern for argmax approach applies to general ML models, and we first outline the concern generically. We later then apply this to the specific setting of image cropping. We term this impact as \textit{argmax bias}.

\begin{figure}
    \centering
    \includegraphics[width=\textwidth]{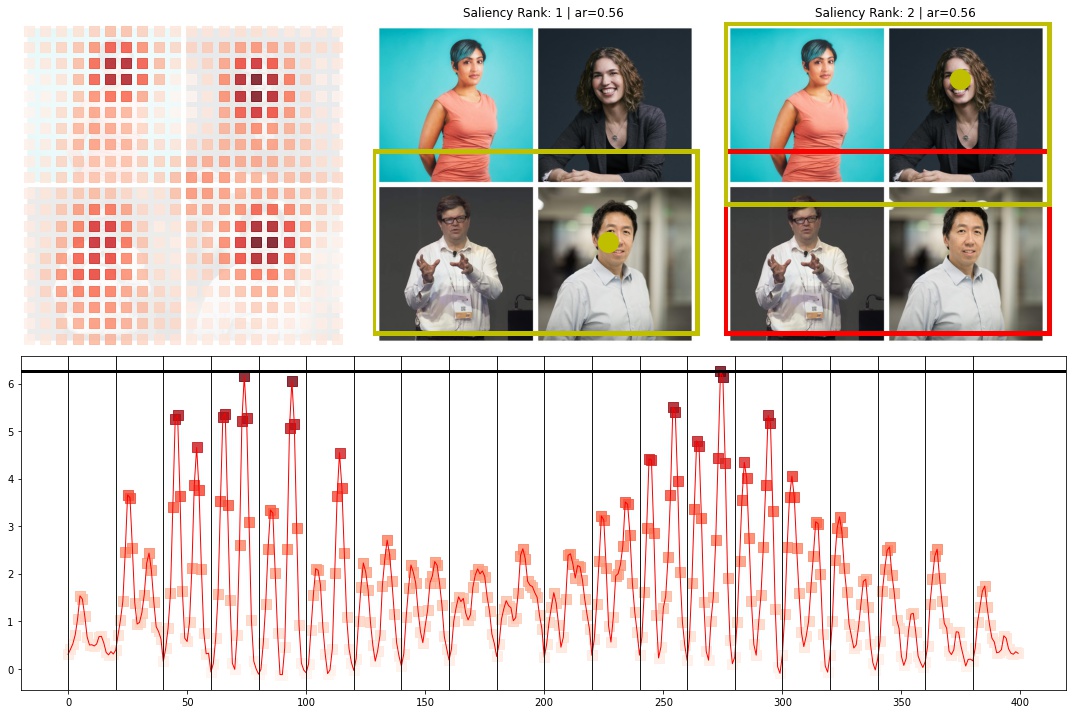}
    \caption{An image with multiple AI researchers (source: \cite{venturebeat}) with multiple almost equally likely salient points. \textbf{Top:} the heat map and the crops based on the first and second most salient points. \textbf{Bottom:} saliency scores of all salient points in the images (top left to bottom right, each row is demarcated by vertical slices). Always cropping based on the top salient point amplifies this crop on the social system compared to the next best salient point.}
    \label{fig:multimodal-crops}
\end{figure}

\paragraph{In General ML}
Given that many machine learning models are probabilistic (i.e. they model $p(y | x)$ for every possible output $y$ for a given input $x$), they should be utilized probabilistically as well, i.e. for inference one should sample $y$ from their output distribution $p(y | x)$ as opposed to always using the most probable $y$. Additionally, in general setting using the most probable $y$ suffices as the this gives the least error when $x$ is sampled from $p(x)$ which is the training data distribution. However, a serious concern arises when $y$ have social impact, e.g. choosing one person over the other in prediction. Disparate impact is compounded when these outputs enter a social system where the outputs have an outreach which follows the power-law distribution, and as a consequence make the chosen $y$ appear as a ground truth for the given $x$ simply because of the deterministic inference system. The agents in the social system do not see all possible realizations of the outputs from $p(y | x)$ and can assume false facts, or "stereotypes" of $x$. More research using user studies can be conducted to validate this theory. For popular images where one individual is highlighted more, we can expect to experience the rich get richer phenomenon where the audience of the platform will identify the selected individual with the image (and its associated event) more than other equally deserving individuals present in the image. This amplification of model predictions in power-law distributed data is what we call the \textit{argmax bias} of the model. A precise quantification of this argmax bias will be the focus of a follow up work.
\paragraph{In Image Cropping}
In our image cropping setting, this issue manifests whenever the image has multiple almost equally likely salient points 
(see Figure \ref{fig:multimodal-crops}). 
This in fact occurs in images with multiple individuals, as the saliency scores on different individuals usually differ by a slight value. The usage of argmax limits only a single point to be selected, giving the impression that all the other points are not salient. In the case of social media platform, if multiple users upload the same image, all of them will see the same crop corresponding to the most salient point. This effect is compounded in the presence of a power law distribution of the shares of images, where certain images, either controversial or from popular accounts, are re-shared across the platform in a disproportionate amount compared to other images, thereby amplifying this effect. As we can see in Figure \ref{fig:multimodal-crops}, selecting the second-best salient point shifts the crop to the top part of the image from the bottom part, highlighting the argmax limitation in presence of competing scores. Also, each slice of the bottom plot in Figure \ref{fig:multimodal-crops} represents a row of saliency score in the images, and we observe almost similar scores for top and bottom halves of the image. Furthermore, the argmax limitation to express more than one salient region still applies even if the competing scores are exactly equal, as the model can only select one point, and as long as the tie-breaking algorithm is deterministic, the selected point will remain the same. This argmax limitation suggests a possible fix based on sampling from the model or allowing the user to select among possible salient regions the best ones. We further discuss this in Section \ref{sec:solutions}. 

\subsection{Other Possible Contributions to the Disparate Impact} \label{sec:other_explanation_bias}
One possible explanation of disparate impact is that the model favors high contrast, which is more strongly present in lighter skin on dark background or darker eyes on lighter skin, and in female heads which have higher image variability \cite{russell2009sex} (see below). Following on the work of \citet{benjamin} who describes how photography and image processing have historically been developed to accommodate whiteness, despite seemingly "neutral" explanations for disparate impact such as lighting and contrast, we emphasize that any plausible explanations related to contrast or variability do not excuse disparate impact. The following observations give some support for this explanation, but they are speculative and not yet conclusive. Limitations of this subsection can be found in Section \ref{sec:limit-explain}. 
 
\paragraph{Dark Backgrounds Contrast Light Skin Colors} 
First, the image attaching affects the saliency prediction (Figure \ref{fig:result_no_attach} vs Figure \ref{fig:base_result}), implying that the model \textbf{does} look at the overall images before assigning saliency scores. We manually observe that majority of images in WikiCeleb have dark backgrounds, 
suggesting that the model's favor for white people may be explained by lighter skin tone having higher contrast to overall images. 
\footnote{This potential explanation is also in line with an informal experiment \cite{vinayarticle} using white and Black individuals on the plain white background. In this case, the favor is observed for Black over white, which is flipped as expected as the background is now white.}

\paragraph{Dark Eyes Contrast Light Skin Tone} On several occasions the most salient point is at the person's eye. We manually observe that eyes' color are mostly on darker side, suggesting that the model's favor for lighter skin tone is a result of stronger contrast between darker eyes with lighter skin surrounding it.
\footnote{Another preliminary results that may support this explanation are the results when we compare whites and Blacks with Asians. Asians are more favored than Blacks and whites with probability $0.62\pm0.04$ and $0.60\pm0.04$, respectively. The fact that Asians are more favored than whites may be due to higher contrast in Asian eyes: Asians and whites do not differ substantially on skin tone, but Asians have significantly darker eyes overall. However, we do not have a large number of samples of Asian images, and hence we omit this result from the main body of the paper.}

\begin{figure}[!tbp]
  \centering
  \includegraphics[width=0.49\textwidth]{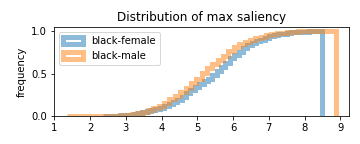}
  \includegraphics[width=0.49\textwidth]{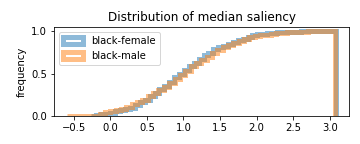}
  \includegraphics[width=0.49\textwidth]{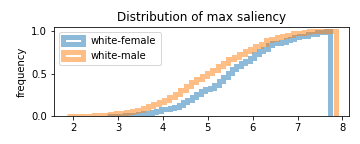}
  \includegraphics[width=0.49\textwidth]{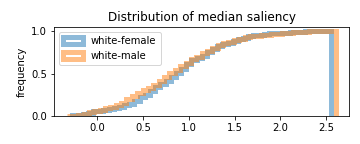}
  \caption{Empirical cumulative distribution functions (ECDFs) of maximum (left) and median (right) saliency scores across all images of Black (top) and White (bottom) individuals, separated by gender. Curves of ECDFs that are on the left are the distributions that on overall take smaller values than ones on the right. The wider the gap, the larger the differences of values.}
  \label{fig:dist_salient}
\end{figure}

\paragraph{Higher Variability in Female than Male Images}
Female heads have higher variability than male heads, which may be party attributed to make-up being more commonly used by females than males \cite{russell2009sex}. This observation is consistent with the explanation that higher contrast increases saliency scores and with the observed favor for female over male.
\paragraph{Higher Variability of Saliency Scores in Females than Male Images} For each image on each subgroup, we run the saliency model and record its  maximum and median saliency score. We then aggregate these statistics for each subgroup, and plot their histograms in Figure \ref{fig:dist_salient}. We found that disparate impact across gender corresponds to histograms of the maximum salient score on female skew to the right compared to male. However, the median histograms on both genders do not show the same skew. Thus, higher maximum saliency in subgroup comes from higher variability in saliency score, but not the overall saliency. Thus, higher variability in female images may lead to higher variability in saliency scores, contributing to the disparate impact in the results.

While the discussions are primarily on Twitter's saliency model, the observations on contrasts and difference in variability of male and female images are educative also for any image model that is sensitive to contrasts. For example, an image model relying on saliency that seeks to detect an important region to create a caption could potentially be impacted on fairness due to these observations as well.

\section{A Qualitative Critique of Machine Learning Based Image Cropping and the Limits of Demographic Parity in Surfacing Representational Harm}
\label{sec:qualitativecriticism}
In this section, we answer the third research question on other important aspects besides systematic disparate impact for consideration in designing image cropping, with the intent of situating risk of harm from image cropping within the broader history of representational harms against marginalized groups, especially Black people and women, who were the focus of our quantitative analysis. Our analysis thus far has revealed systematic disparities in the behavior of the image cropping model under the notion of demographic parity, which is meant to alleviate concerns of under-representation. However, even in the ideal world where the model exhibits zero problems with regards to demographic parity under this framework, there are deeper concerns. Although the analysis above was able to demonstrate some aspects of model behavior leading to disparate impact, including unforeseen consequences of using argmax and the potential effect of contrast for saliency-based cropping, there are inherent limitations of formalized metrics for fairness. They are unable to sufficiently elucidate all the relevant harms related to automatic image cropping.

The primary concern associated with automated cropping is representational harm, which is multi-faceted, culturally situated \cite{crawford2017trouble, benjamin}, and difficult to quantify \cite{crawford2017trouble}. The choice of who to crop is laden with highly contextual and nuanced social ramifications that the model cannot understand, and that the analysis using demographic parity cannot surface. For example, think of how the interpretation of cropping could differ between a photo of two people, one darker-skinned and one lighter-skinned, if the context is they are software engineers versus criminals. Unfortunately, the interpretation of such images occurs through \citet{butler2020endangered}'s racially saturated field of visibility, where dominant conceptions of race may affect viewers interpretation of the image. 
Similarly, in comparing the hyper-sexualization and over-representation of Black women in the pornography industry with the lack of Black female nudes in the Western art canon, which \citet{nash2008strange} argues represents the lack of art that celebrates the beauty and femininity of Black woman, \citet{nash2008strange} calls for ``conceptualizing both visibility and invisibility in more historically contingent and specific terms.'' Although demographic parity can be a useful signal for surfacing systematic differences in treatment, on its own it becomes a crude and insufficient metric for addressing under-representation when viewed through this lens. We recognize the inherent limitations of formalized metrics for fairness, which are unable to capture the historically specific and culturally contextual meaning inscribed unto images when they are posted on social media.  For marginalized groups, clumsy portrayals and missteps that lead to the removal of positive in-group representation are especially painful since there are frequently very few other positive representations to supplement, which unfortunately places disproportionate emphasis on the few surviving positive examples \cite{disclosure}.

Similar limitations exist for the formal analysis attempting to address issues of male gaze.  Although we find no evidence the saliency model explicitly encodes male gaze, in cases where the model crops out a woman's head in favor of their body due to a jersey or lettering, the chosen crop still runs the risk of representational harm for women. In these instances, it is important to remember that users are unaware of the details of the saliency model and the cropping algorithm. Regardless of the underlying mechanism, when an image cropped to a woman’s body area is viewed on social media, due to the historical hyper-sexualization and objectification of women's bodies \cite{mackinnon1987feminism, fredrickson1997objectification}, the image runs this risk of being instilled with the notion of male gaze. This risk is, of course, not universal and context dependent; however, it underscores the fact that the underlying problem with automatic cropping is that it cannot understand such context and it should not be dictating so strongly the presentation of women's bodies on the platform. Framing concerns about automated image cropping purely in terms of demographic parity fails to question the normative assumption of saliency-based cropping: the notion that for any image, there is a "best"--or at the very least "acceptable"--crop that can be predicted based on human eye tracking data. This critique echos previous calls to question machine learning's imposition of a normative "optimal state" in all aspects of life, especially those related to human expression and the arts \cite{vallor_2021}. Machine learning based cropping is fundamentally flawed because it removes user agency and restricts user’s expression of their own identity and values, instead imposing a normative gaze about which part of the image is considered the most interesting. Posing concerns related to image cropping purely in terms of demographic parity fails to question the "closure of predetermined,technical problems" \cite{gasson2003human} and fails to consider sociotechnical considerations, an example of the "framing trap" \cite{selbst2019fairness}.
Allowing users more control over how their photos are presented better accommodates pluralistic design practices \cite{escobar2018designs,tasioulas_2021,vallor_2021}.

Twitter is an example of an environment where the risk of representational harm is high since Twitter is used to discuss social issues and sensitive subject matter. In addition, Tweets can be potentially viewed by millions of people, meaning mistakes can have far reaching impact. 
Given this context\footnote{Automatic image cropping could be more appropriate in other more limited contexts that deal with less sensitive content, smaller audiences, less variability in the types of photos cropped, etc.}, we acknowledge that automatic cropping based on machine learning is unable to provide a satisfactory solution that can fully address concerns about representational harm and user agency since they place unnecessary limits on the identities and representations users choose for themselves when they post images. Simply aiming for demographic parity ignores the nuanced and complex history of representational harm and the importance of user agency.

\section{Recommendations} \label{sec:solutions}
%--------------------------------------------------
% Define environment for using itemize within tabularx

\newenvironment{tableitems}{\begin{itemize}[noitemsep,topsep=0pt,leftmargin=*,after=\vspace*{-\dimexpr\baselineskip},before=\vspace*{-0.5\baselineskip}]}{\end{itemize}}

\begin{table}
  \caption{A non-exhaustive comparison of approaches to implement Image Cropping and their salient properties and details. Hybrid solutions among these are also possible. \textbf{User Agency} refers to the amount of control the user has on how their image is represented on the platform. \textbf{Automation} refers to the ability to automate the cropping process. \textbf{Coverage} refers to what proportion of images whose crops are satisfactory to users. \textbf{Risk on multi-modal images} refers to the risk of disparate impact when images contain multiple highly salient points, i.e. risk of the argmax bias (Section \ref{sec:argmax}).}
  \label{tab:solutions}
  \small % for final version, the font is large and makes the table overfit the page
  \begin{tabularx}{\textwidth}{>{\raggedright\arraybackslash}p{0.23\textwidth}>{\raggedright\arraybackslash}p{0.25\textwidth}>{\raggedright\arraybackslash}p{0.47\textwidth}}
    \toprule
    \textbf{Approaches} & \textbf{Properties} & \textbf{Comments}\\
    \midrule
    \textbf{Argmax saliency-based cropping (Twitter model)}        
        &   \begin{tableitems}
            \item \textbf{User Agency}: None
            \item \textbf{Automation}: Full
            \item \textbf{Coverage}: Majority
            \item \textbf{Risk on multi-modal images}: High
            \end{tableitems}
        &   \begin{tableitems}
        \item Suffer ``argmax bias'': small difference in saliency scores can easily lead to systematic discrepancies
        \item The coverage is majority of images on a light assumption that the majority of images have a single highly salient region.
        \end{tableitems}
            \\
            \midrule
    \textbf{Sampling saliency-based cropping}: sample the focal point with probability equal predicted saliency scores              
        &   \begin{tableitems}
            \item \textbf{User Agency}: None
            \item \textbf{Automation}: Full
            \item \textbf{Coverage}: High
            \item \textbf{Risk on multi-modal images}: Reduced
            \end{tableitems}
        &   \begin{tableitems}
            \item Cropping can vary greatly even when images are the same or vary very little, giving unpredictable behavior to users
            \item There is some probability from sampling that a crop focuses on a low salient region, reducing the coverage.
            \end{tableitems}
        \\ \midrule
    \textbf{Averaging saliency-based cropping}: use the average weighted by predicted saliency scores or average over top $k$ salient points as the focal point              
        &   \begin{tableitems}
            \item \textbf{User Agency}: None
            \item \textbf{Automation}: Full
            \item \textbf{Coverage}: High
            \item \textbf{Risk on multi-modal images}: Reduced
            \end{tableitems}
        &   \begin{tableitems}
            \item Less stable outputs. Crops can move by small edits, e.g. an additional text at the bottom
            \item Crops can focus on a low salient region, e.g. focus at the middle when two salient regions are on the left and right
            \end{tableitems}
        \\ \midrule
    \textbf{User choice among $k$ points}: Providing $k$ most salient points and let the user pick one of those points to crop around
        &   \begin{tableitems}
            \item \textbf{User Agency}: Some
            \item \textbf{Automation}: Some
            \item \textbf{Coverage}: Higher
            \item \textbf{Risk on multi-modal images}: Greatly Reduced
            \end{tableitems}
        &   \begin{tableitems}
            \item Users have more burden as $k$ increases
            \item Some heuristics or studies are required to ensure top $k$ points represent the images well, e.g. not being too clustered in one area, and to specify $k$ effectively
            \item Shift the problem of picking most salient point among many highly salient points to users
            \item There is still risk (which is smaller as $k$ increases) that no top $k$ points are satisfactory
            \end{tableitems}
        \\ \midrule
    \textbf{Focal point selection}: let the user pick the focal point to crop around                   &   
        \begin{tableitems}
        \item \textbf{User Agency}: Full
        \item \textbf{Automation}: None
        \item \textbf{Coverage}: Full
        \item \textbf{Risk on multi-modal images}: None

        \end{tableitems}
        &   \begin{tableitems}
            \item One action of user participation is enough to define cropping on multiple aspect ratios        
            \item Give users cognitive load to select the point. Can be bothersome to users who are not concerned about their crops
            \item Shift the problem of picking most salient point among many highly salient points to users
            \end{tableitems}
        \\ \midrule
    \textbf{No crop or pad-to-fit}: don't crop image or pad it to fit desired aspect ratios 
        &   \begin{tableitems}
            \item \textbf{User Agency}: WYSIWYG (what you see is what you get, i.e. the image is represented as-is)
            \item \textbf{Automation}: Full
            \item \textbf{Coverage}: Full
            \item \textbf{Risk on multi-modal images}: None
            \end{tableitems}
        &   \begin{tableitems}
            \item Easy to implement; no prediction model required
            \item Images may contain padded background which may reduce the effective use and the attractiveness of the platform space
            \item Images with very high or low aspect ratios will be scaled to a very small size; fitted images are more likely to be unreadable
            \item Because users can expect their images to be presented the same way they upload them, they have full control over the cropping of the image
            \end{tableitems}
        \\
  \bottomrule
\end{tabularx}
\end{table}

In this section, we answer the fourth research question on alternatives to Twitter saliency-based image cropping.
In order to sufficiently mitigate the risk of representational harm on Twitter, ML-based image cropping may be replaced in favor of a solution that better preserves user agency. Removing cropping where possible and displaying the original image is ideal, although photos that aren’t a standard size which are very long or wide pose challenging edge cases. Solutions that preserve the creator’s intended focal point without requiring the user to tediously define a crop for each aspect ratio for each photo are desirable. In Table \ref{tab:solutions}, we present a list of solutions. The list is non-exhaustive, and a solution combining multiple approaches is also possible. We consider possible solutions along the following dimensions: \textbf{User Agency}, \textbf{Automation}, \textbf{Coverage}, and \textbf{Risk on multi-modal images}. See the Table's caption and comments for more details.

In evaluating tradeoffs between solutions, we observe the same tensions raised in early debates about weighing the value of user control against the value of saving user's time and attention by automating decisions \cite{shneiderman1997direct}, although these two approaches are not necessarily mutually exclusive \cite{dove2017ux}. Solutions that amplify the productivity of users by decreasing the amount of attention needed to generate satisfactory crops as much as possible while maintaining a sense of user control are ideal \cite{shneiderman1997direct}. 
There can also be hybrid among these solutions. For example, we perform saliency-based cropping, but ask the users to confirm the crop if the top few saliency points spread in the picture, and allow users to specify the focal point if they wish. 
 Each of these solutions may present their own trade-offs compared to the original fully-automated image cropping solution and their exact utility 
 and trade-offs require further investigation via a user study. 
In order to properly assess the potential harms of technology and develop a novel solution, our analysis reaffirms the following design implications that have been previously discussed within the ML and HCI communities:
 \begin{enumerate}
    \item \textbf{Co-construction of the core research activities and
value-oriented goals \cite{bardzell, friedman2002value, friedman2008value}} 
   
    The high-level goal of this work is to understand the societal effects of automatic cropping and ensure that it does not reproduce the subordination of marginalized groups, which has informed our recommendation to remove saliency-based cropping and influenced our decision to leverage group fairness metrics in combination with a qualitative critique.
      \item \textbf{The utility of combining qualitative and quantitative methods \cite{bardzell, friedman2002value, friedman2008value}} 
    
    As argued in Section \ref{sec:qualitativecriticism}, although it is tempting to view concerns about image cropping purely in terms of group fairness metrics, such an analysis in isolation does not adequately address concerns of representational harm.  
     In addition, in Section \ref{sec:limit_and_future_work}, we also draw on previous work in sociology, feminist theory, and critical race theory in discussing the positionality, limitations and risks of using standardized racial and gender taxonomies. The importance of user agency in evaluating solutions is also motivated by the notion that self-representation is an important tool for dismantling dominant representations and stereotypes \cite{collins2002black}.

    \item \textbf{The importance of centering the experience of marginalized peoples \cite{hanna2020towards,bardzell,costanza2018design}} Understanding the potential severity of representational harm requires an understanding of how representational harm has historically impacted marginalized communities and how it reinforces systems of subordination that lead to allocative harm \cite{crawford2017trouble, sweeney2013discrimination, noble}, as discussed in Section \ref{sec:historyrepresentationalharm}. 
    
     \item \textbf{Increased collaboration between ML practitioners and designers in developing ethical technology \cite{yang2017role}}
     
    We observe that saliency based cropping presupposes a universally "best" crop, implicitly imposing a normativity that does not accommodate the values of pluralism and self-determination. Our critique motivates an emphasis on user control and human-centered design to better understand the values implicitly embedded in design decisions. These have been long standing topics of research within the HCI community \cite{shneiderman1997direct, borning2012next, friedman2002value, friedman2008value, bardzell, costanza2018design}. In order to properly evaluate the trade-offs of the various solutions presented in Table \ref{tab:solutions} and develop a viable alternative, we recommend additional user studies, with a focus on understanding the concerns and perspectives of users from marginalized communities.
    
       \item \textbf{In developing ethical technologies, moving from a fairness/bias framing to a discussion of harms}
       
     As we have illustrated here with respect to automatic image cropping and as other have noted \cite{barocas-hardt-narayanan,corbett2018measure}, formalized notions of fairness often fail to surface all potential harms in the deployment of technology. The fairness framing is often too narrowly scoped as it only considers difference of treatment within the algorithm and fails to incorporate an holistic understanding of the socio-technical system in which the technology is deployed \cite{selbst2019fairness, kasy2021fairness}. By centering our analysis on a discussion of potential harms, we are better equipped to address issues of stereotyping and male gaze, both of which are highly dependent on historical context.
    
 \end{enumerate}

 Human centered and ethical design frameworks provide useful tools to understand the normative values embedded in design, and evaluating who is afforded or dysafforded by a design decision \cite{costanza2018design}. In evaluating alternative solutions, our analysis motivates further user studies and input from designers to better understand users' concerns and proactively elucidate the values embedded in design and potential harms.
 
\subsection{Twitter's Product Changes}
Twitter has committed to product changes after users raised concerns on the platform about its image cropping algorithm.
In October 2020, Twitter committed to making product changes to reduce its reliance on the image cropping algorithm, instead prioritizing user agency. In a blog post, the company said, ``giving people more choices for image cropping and previewing what they’ll look like in the Tweet composer may help reduce the risk of harm'' \cite{twitterresponse}.
In March 2021, one image cropping solution was released by Twitter\citeTweet{1369682375668424707}
to a small group of users, and was later launched for all users in May 2021\citeTweet{1390040111228723200}. The solution was similar to the "No cropping" alternative in Table 1, and it includes the removal of the saliency algorithm for image cropping. The change was motivated to better align with people’s expectations: ``how to crop an image is a decision best made by people'', and ``the goal of this [change] was to give people more control over how their images appear while also improving the experience of people seeing the images in their timeline'' \cite{twitterresponse2}.

\subsection{Recommendations Beyond Image Cropping}

The recommended solutions are motivated from two concerns: the argmax approach (Section \ref{sec:argmax}) and the lack of user agency creating representational harms (Section \ref{sec:qualitativecriticism}).
\paragraph{Addressing the argmax approach.} The first two alternatives in Table \ref{tab:solutions}, namely sampling and averaging saliency-based cropping, aim to eliminate amplified disparate impact due to the argmax approach, even though they can have more sensitivity to input than using the argmax function. As argmax bias can be a concern for any general ML models, these alternatives are also options for general ML practitioners to implement to potentially reduce disparate impact. Sampling from or averaging the top-$k$ predictions may be a suitable intervention to alleviate argmax bias for a variety of ML problems.
\paragraph{Insufficiency of Demographic Parity in Surfacing Risk of Representational Harm} Although demographic parity can be a useful tool for measuring systematic disparities in treatment, because of the highly contextual nature of representational harms, demographic parity on its own is insufficient for surfacing representational harms such as stereotyping and under-representation. This insight is generalizable to any system where representational harms may be an issue.
\paragraph{Addressing the lack of user agency.} The last three alternatives in Table \ref{tab:solutions} aim to reduce representational harms. They include user participation at different levels, from none to high involvement. Therefore, they can be similarly used in other settings where there are concerns of representational harms. First, \textit{User choice among $k$ points}, can help reduce the argmax bias amplification. Second, \textit{Focal point selection} can be generalized to users provided the label for their content as opposed to models guessing them, e.g. users highlight the topic of their document or key entities in the document as opposed to platforms guessing them. Finally, \textit{No crop} can be generalized to a model-free approach to solving a problem, as sometimes using a machine learning model is not appropriate for a given task if the potential risks are known.

\section{Limitations and Future Directions} 
\label{sec:limit_and_future_work}
In this section, we discuss limitations of our work and potential future work to address some of those limitations.

\paragraph{Race and Gender Labels and WikiCeleb dataset} Standardized racial and gender labels can be too limiting to how a person wants to be represented and do not capture the nuances of race and gender; for example, it is potentially problematic or even disrespectful for mixed-race or non-binary gender individuals. Additionally, the conceptualization of race as a fixed attribute poses challenges due to its inconsistent conceptualization across disciplines and its multidimensionality \cite{hanna2020towards, scheuerman2020we}; these problems are especially relevant given we are using labels from Wikidata (Section \ref{sec:dataset}), which is curated by a diverse group of people.  Given ethnic group labels from Wikidata, we refer to the US census race categories for how to standardize and simplify these categories into light- and dark-skinned in order to conduct our analysis.

Following the suggestions in \cite{scheuerman2020we}, we also wish to give a brief description of the sociohistorical context and positionality of our racial and gender annotations to better elucidate their limitations.
Wikidata primarily features celebrities from the US and the Western world. Using Wikidata images and the US census as a reference presents a very US-centric conception of race.  Many critical race scholars also define race in terms of marked or perceived difference, rather than traceable ancestry in itself \cite{haslanger2000gender,hirschman2004origins,scheuerman2020we}. Additionally, people of shared ancestry or the same racial identity can still look very different, making racial categories based on ancestry not always a suitable attribute to relate to images.  

For future directions, one alternative to racial identity is skin tone, such as in \citet{buolamwini2018}, which is more directly related to the color in the images.\footnote{We were able to use the GenderShade dataset \cite{buolamwini2018} at the time of publication due to a licensing issue.}  
Using a more fine grained racial and gender taxonomy may alleviate some concerns related to the black/white binary \cite{perea1998black} and the gender binary, although one challenge related to using more fine grained labels is the need for sufficient sample sizes.
 In addition, a more fundamental critique of racial and gender taxonomies is that they pose a risk of reifying racial and gender categories as natural and inherent rather than socially constructed \cite{hanna2020towards,fields2014racecraft, benthall2019racial}. The use of standardized racial and gender taxonomy here is not meant to essentialize or naturalize the construction of race and gender, or to imply that race is solely dependent on ancestry, but rather to study the impact of representational harm on historically marginalized populations \cite{noble,hanna2020towards}. 

\paragraph{Plausible Explanations} \label{sec:limit-explain}

The plausible explanations for disparate impact in Section \ref{sec:other_explanation_bias} are only suggestive and not conclusive, and there are other possible underlying causes. An example is that the maximum saliency scores across subgroups (Figure \ref{fig:dist_salient}), whose disparities result in systematic disparate impact, are a combination of facial and background regions of images. While maximum saliency points usually lie on heads, on occasions they are on graphics such as texts or company logos on backgrounds or clothes, such as when we discuss male gaze in Section \ref{male gaze section}. These small subsets of images may seem to be favored by the model due to its demographic, but in reality it could be due to their backgrounds. These non-facial salient regions may also contribute to the difference in media saliency scores as shown in Figure \ref{fig:dist_salient}. In addition, we did not study whether human gaze annotations in the training datasets themselves are a contributing factor in the model's disparate impact. 

More experiments and understanding in explainable ML are needed to evaluate relevant factors in the datasets that potentially contribute to model disparity, and to more clearly explain the observed systematic disparate impact. For example, bigger and more diverse datasets with closer inspections on the correlation between the claimed contributors of bias and maximum saliency points can expand this line of research and are the future directions of this work. This is an important aspect as bias related work can often be better explained away by more systematic factors in presence of higher quality datasets and more robust analysis (see \cite{Mishra2018SelfCite}). 

\paragraph{Recommended Solutions}
The pros and cons in recommended solutions in Section \ref{sec:solutions} are provided as starting points to consider. More user and design studies are needed to identify the best cropping methods or some hybrid among those, and other details in the chosen method such as selecting the parameter $k$ if needed.

\section{Conclusion} \label{sec:conclusion}
Twitter's saliency-based image cropping algorithm automatically crops images to different aspects ratios by centering crops around the most salient area, the area predicted to hold human's gaze. The use of this model poses concerns that Twitter's cropping system favors cropping light-skinned over dark-skinned individuals and favors cropping woman's bodies over their heads. At the first glance, it may seem that the risk of harm from automated image cropping can be articulated purely as a fairness concern that can be quantified using formalized fairness metrics. We perform a fairness analysis to evaluate demographic parity (or the lack thereof, i.e. disparate impact) of the model across race and gender. We observe disparate impact and outline possible contributing factors. Most notably, cropping based on the single most salient point can amplify small disparities across subgroups due to argmax bias.  

However, there are limitations in using formalized fairness metrics in assessing the harms of technologies. Regardless of the statistical results of fairness analysis, the model presents a risk of representational harm, where the users do not have the choice to represent themselves as intended. Because representational harm is historically specific and culturally contextual, formalized fairness metrics such as demographic parity are insufficient on their own in surfacing potential harms related to automatic image cropping. For example, even if Twitter's cropping system is not systematically favoring cropping women's bodies, concerns of male gaze persist due to the  historical hyper-sexualization and objectification of women's bodies, which influence the interpretation of images when posted on social media. We enumerate alternative approaches to saliency-based image cropping and discuss possible trade-offs. Our analysis motivates  combination of quantitative and qualitative methods that include human-centered design and user experience research in evaluating alternative approaches.

%% UNHIDE DURING CAMERA READY

\section*{Acknowledgement}
We want to thank Luca Belli, Jose Caballero, Rumman Chowdhury, Neal Cohen, Moritz Hardt, Ferenc Huszar, Ariadna Font Llitjós, Nick Matheson, Umashanthi Pavalanathan, and Jutta Williams for reviewing the paper. 

\bibliographystyle{ACM-Reference-Format}
\bibliography{base}


\end{document}